# Premelting fluctuations

Pavel Golovinski

*Physics Research Laboratory, Voronezh State Technical University, Voronezh, Russia*

golovinski@bk.ru

A model of the premelting fluctuations is proposed, based on the Landau mean field theory applied to a first-order phase transition. Using the thermodynamic potential, the nonlinear Langevin equation for the order parameter is formulated, which describes the dynamics of the phase transition considering microscopic thermal fluctuations. The origin of the low-frequency fluctuations, associated with a switching in a bistable system under the influence of a thermal noise, is shown. Analytical two-scale model of the premelting fluctuations is developed, with characteristic frequent small-amplitude fluctuations of chaotic motion in the vicinity of potential minima and rarer large fluctuations due to solid-liquid transitions. Based on the numerical simulation of the solutions to the stochastic differential equation, both the dynamics of the order parameter fluctuations and their spectrum are obtained. The model qualitatively reproduces experimental spectrum of isothermal fluctuations.

## 1. Introduction

Melting is important in a huge variety of natural and artificial phenomena, including the formation of nanostructures. From the thermodynamic point of view, solid and liquid phases coexist during melting, and there is no continuous transition between them, i.e. a first-order phase transition occurs abruptly, and the dependence of the order parameter on temperature is discontinuous. On the other hand, if you look at melting from a microscopic point of view, then near the melting temperature you can trace the state of melting with partial disordering in the bulk crystal and a continuous transition layer between the solid and liquid state at the surface [1]. A theoretical description of surface melting was undertaken in the classical phonon model, lattice models, the Landau phenomenological theory, and in the Kosterlitz-Thoules approximation, which appeals to the thermal excitation of dislocations [2]. In addition, the theory of the density functional [3] and computer simulation based on the molecular dynamics method [4-6] were used. Direct experiments [7] with proton scattering by the surface of heated lead confirmed that surface melting is started below the bulk melting temperature.

The mechanism of formation a liquid film on the surface of a solid during incomplete melting with prediction of its thickness is described by the Dzyaloshinsky-Lifshits-Pitaevsky theory, which takes into account the dispersion forces of interaction in a medium [8]. A statistical consideration of this layer is performed also in the lattice melting theory [9]. In this way, despite the fact that fluctuations were not considered, a decreasing dependence of the density on the thickness of the liquid on the solid surface is obtained. Incomplete melting, in addition, depends on the orientation of the crystal lattice with respect to the surface [10], and when melting nanosize particles, the boundary between surface melting and melting of the entire sample is absent, with the critical temperature depended of the particle sizes [11]. The transition from bulk to surface melting occurs at particle sizes of about 10 nm depending on a substance [12, 13].

The variety of observed effects and theoretical speculations in the field is predetermined unflagged research interest in a deeper understanding of premelting. In [14], the pre- and post-melting of the crystalline substances were studied by differential thermal analysis, and slow fluctuations with frequencies in the range 0.04–0.25 Hz were detected. In the reverse crystallization process, these phenomena were not observed. Later [15], the method of digital differential thermal analysis was used to study in detail the spectrum of fluctuations in the heat flow for Cu, Sb, Ge, and KCl crystals under isothermal premelting conditions, which demonstrates a universal two-level structure of flicker noise with different behavior in the high-frequency and low-frequency domains.

In our work, premelting fluctuations are considered using the nonlinear Langevin equation for order parameter. The article is structured as follows. In Section 2, the main principles of the mean field theory for the phase transitions of a first kind are formulated. In Section 3, the problem of fluctuations in premelting is reduced to solving the Langevin equation for a bistable oscillator. Based on the two-scale model, an analytical expression is obtained for the spectrum of fluctuations of premelting. In Section 4, a numerical simulation of the nonlinear Langevin equation is performed. The obtained theoretical frequency spectrum is compared with experimental data and results of calculations in the analytical model.

## 2. The mean field approximation for first-order phase transitions

Assume that at first, the crystal is in equilibrium with the surrounding vapor. As the temperature $T$ gradually rises to the critical point, the bulk crystal remains almost unchanged up to the melting point $T_m$ when a first-order phase transition is observed. First-order phase transitions can be considered using the Gibbs free energy expansion in a Taylor series in terms of order parameter $\eta$ [16]:

$$G(\eta,T) = G(0,T) + a_1\eta + \frac{a_2}{2}\eta^2 + \frac{a_3}{3}\eta^3 + \frac{a_4}{4}\eta^4 + ... , \qquad (1)$$

representing it in the form of a power law potential. The order parameter here is the relative difference in the density of the substance in the solid and liquid phase.

For the phase transitions of a first kind we have parameters $a_1 = 0$, $a_3 < 0, a_4 \neq 0$, $a_2 = \alpha(T - T_c)$, and, while keeping the minimum number of the most important terms, the Gibbs potential takes the form

$$G(\eta,T) = G(0,T) + \frac{\alpha(T-T_c)}{2}\eta^2 - \frac{|a_3|}{3}\eta^3 + \frac{a_4}{4}\eta^4 . \qquad (2)$$

In the expression for free energy, in contrast to the case of second-order phase transitions, there is a third power of the order parameter [17]. This property allows us to describe the phase transition, which is discontinuous, i.e. occurs abruptly. The model demonstrates typical features of a first-order transition, but it has no universal applicability as $\eta$ may not be small, unlike the case for continuous transitions.

A common feature of the phase transitions of a first kind is the coexistence of two phases with simultaneous minima at two points $\eta = 0$ and $\eta \neq 0$, which leads to the appearance of a hysteresis phenomenon. When the temperature changes, the system can remain in a local potential minimum, although it may be not the state with the lowest free energy. The potential barrier separating stable states can be overcome under the influence of random fluctuations always existing in real systems. This explains the observed finite time of conservation the metastable states.

The extrema of the function $G$ are determined by the condition $\partial G / \partial \eta = a_2\eta + a_3\eta^2 + a_4\eta^3 = 0$, which gives values

$$\eta_1 = 0, \quad \eta_{2,3} = -\frac{a_3}{2a_4} \pm \sqrt{\left(\frac{a_3}{2a_4}\right)^2 - \frac{a_2}{a_4}}. \qquad (3)$$

At a temperature $T_m$ for which

$$G(\eta_1, T) = G(\eta_3, T), \qquad (4)$$

a jump in the ordering parameter from $\eta_1$ to $\eta_3$ takes place, which is an evidence of a first-order phase transition [18] (Figure 1). In a disordered state, the order parameter $\eta = 0$, while for the ordered state, $\eta > 0$, which is true for most crystalline substances, except for water and bismuth, which are characterized by an anomalous increase in density during melting.

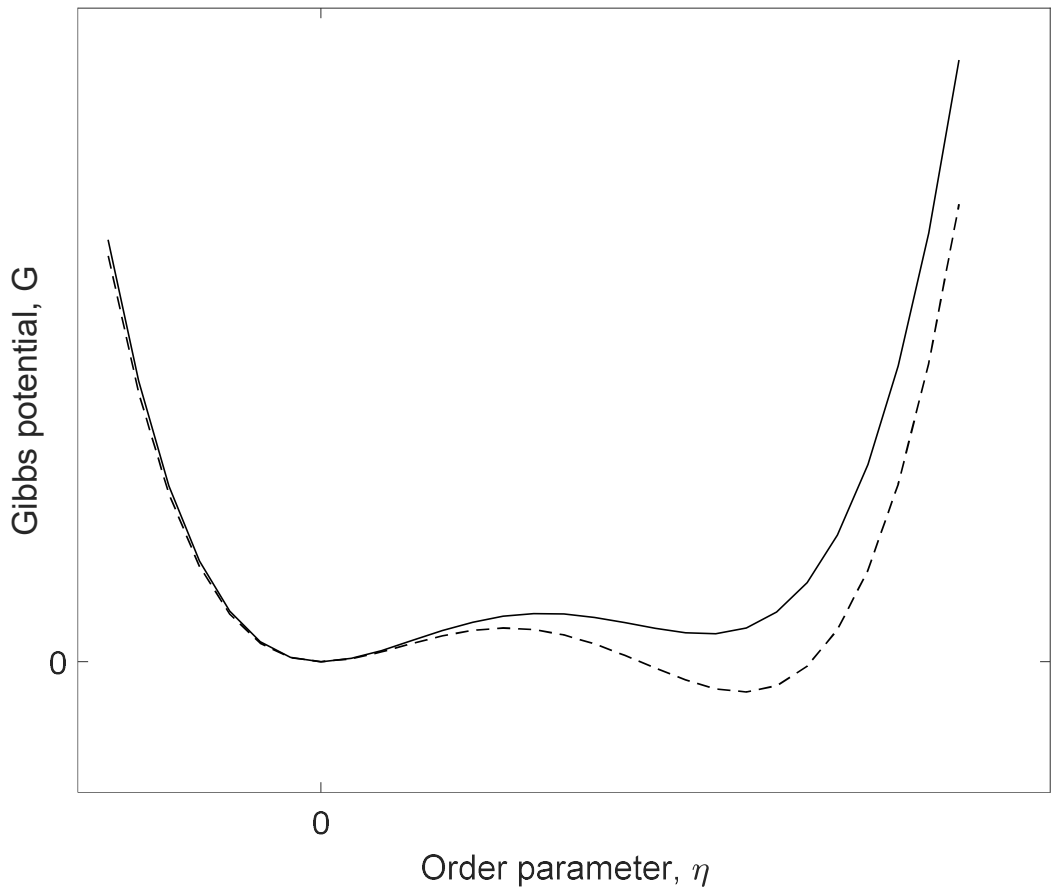

Figure 1. Characteristic dependence of the bulk free energy on the order parameter for a first-order phase transition. The upper solid curve corresponds to $T > T_m$, the lower dashed curve shows the dependence at $T < T_m$.

In the immediate vicinity of the critical point, in the premelting conditions, the depths of the two minima are close, so that, in accordance with Polyakov's general hypothesis about increasing the system symmetry near the phase transition, dependence (2) becomes symmetric with respect to the parameter value $\eta = \eta_2$ [19]. Hence, by choosing the scale and shifting $\eta$, the symmetric bistable potential can be represented in the form

$$G(\eta) = -\eta^2/2 + \eta^4/4, \qquad (5)$$

which is used in further consideration of premelting fluctuations.

## 3. Stochastic dynamics of order parameter

At critical points where the system changes its macroscopic state, fluctuations play a decisive role, and the problem of description bifurcations, being characteristic for the theory of equilibrium states, turns into the problem of the dynamics of nonequilibrium phase transitions. Near the transition points, fluctuations of the order parameter become especially large, and so-called "critical fluctuations" appear. The resulting phase transitions induced by thermal noise can be described based on the theory of random processes using stochastic differential equations [20].

To understand flicker noise experimentally observed in the phase transitions, an explanation based on the mean field theory was proposed [21, 22]. The main idea was to convert a white noise by a nonlinear system in fluctuations with a $1/f$ spectrum. This model assumes coexisting of two phases, each of which is described by its own order parameter. In this model, the flicker noise arises both in spatially homogeneous and spatially inhomogeneous systems, which are described by the modified Ginzburg-Landau potential for two field variables. We propose a modified approach based directly on equation (5) for the one-parameter Gibbs potential of critical bistable state in the phase transitions of a first kind. In such double-well bistable systems under the action of a random force, a complex chaotic dynamics arises, capable of generating spectra with slowly decreasing asymptotic [23]. A bistable system in the absence of noise rests on in one of the stable states. Under the influence of environmental fluctuations, transitions from one state to another occur. Near the point of macroscopic instability, noise-induced fluctuations lead to a macroscopic switching effect. Under such conditions, the fluctuations of the medium can no longer be regarded as a small perturbation [24], and two attractors are formed in a bistable system, with transitions inducing by external noise [25].

The relaxation equation of the order parameter when it deviates from the equilibrium, in its simplest form, without consideration spatial inhomogeneity, can be written as the nonlinear Langevin equation [20, 24]

$$\gamma \frac{\partial \eta}{\partial t} = -\frac{\partial G}{\partial \eta} + F(t), \qquad (6)$$

where the parameter $\gamma$ regulates the rate of relaxation to the stable phase. For further consideration, it is taken equal to unity, by inclusion in the time scale. The term $F(t) = \sqrt{2D}\xi(t)$ defines a random Gaussian white noise with a zero mean value, which is added to consider thermal fluctuations of a microscopic origin. Random noise $F(t)$ belongs to a statistical ensemble with average

$$\langle \xi(t)\xi(t_1) \rangle = \delta(t - t_1), \qquad (7)$$

and the value of the constant $D$ depends on the temperature. Relation (7) indicates that fluctuations are uncorrelated in time. Our model represents fluctuations near the critical point as the equivalent nonlinear Brownian viscous motion of some effective particle.

The chaotic motion of a particle in a potential with two minima is characterized by two different processes: small frequent fluctuations under the action of a random force $F(t)$ near equilibrium position and more rare transitions from one equilibrium position to another [26]. If we restrict ourselves only to transitions between two equilibrium states $\eta_1 = -1$ and $\eta_3 = 1$, then the problem becomes equivalent to a Markov dichotomous noise [27]. In this simplified model, system has only two different states and stays in each of which during average time $\tau_1 = \tau_3$. The equations for the probabilities $P_1$ and $P_3$ to be in one of these states are of the form

$$\frac{dP_1}{dt} = -k_1 P_1 + k_3 P_3, \tag{8}$$

$$\frac{dP_3}{dt} = k_1 P_1 - k_3 P_3,$$

where $k_1 = k_3 = 1/\tau_1$, and the conservation condition of the total probability is $P_1 + P_3 = 1$. In a steady state, we have probabilities $P_1 = P_3 = 1/2$ with a mean value

$$\langle \eta(t) \rangle = 0 \tag{9}$$

and a correlation function

$$\langle \eta(t)\eta(t') \rangle = \frac{D_d}{\tau} \exp\left(-\frac{|t-t'|}{\tau}\right). \tag{10}$$

Here $\tau = \tau_1/2$ is the correlation time, and the amplitude is

$$D_d = \frac{\tau}{2}. \tag{11}$$

The correlation function (10) is Ornstein-Uhlenbeck exponentially correlated color noise. The spectrum of this correlation function has Lorentz shape

$$S_d(\omega) = \frac{D_d}{\pi(1+\omega^2\tau^2)}. \tag{12}$$

To determine parameters $\tau_1$ and $\tau_3$, we follow the Kramers approach [28]. Note that the equation of motion (6) for a first-order phase transition in the critical region takes the form of a time-dependent Ginzburg-Landau equation [20, 29] with a random term

$$\frac{\partial \eta}{\partial t} = \eta - \eta^3 + \sqrt{2D}\xi(t). \tag{13}$$

Then, we represent the Fokker-Planck equation [30] corresponding to equation (13) in the form

$$\frac{\partial P(\eta,t)}{\partial t} = -\frac{\partial}{\partial \eta} J(\eta,t), \tag{14}$$

where the probability flux

$$J(\eta,t) = -\frac{\partial G}{\partial \eta} P(\eta,t) - D \frac{\partial P(\eta,t)}{\partial \eta}. \tag{15}$$

Using the Fokker-Planck equation (14), we can obtain the Kramers formula [31] for the rate of transition over the barrier

$$k_n = \frac{\omega_2 \omega_n}{2\pi} \exp\left(-\frac{\Delta G}{D}\right), \tag{16}$$

where $\omega_n^2 = |G''(\eta_n)| = -1 + 3\eta_n^2$, $n = 1,3$. Under premelting, the Gibbs potential is symmetric and numerically $\omega_1^2 = 2\omega_2^2 = \omega_3^2 = 2$, $\Delta G = 1/4$. So, we determined the correlation time $\tau = (2k_n)^{-1}$ in the dichotomous Markov model and established a relationship between the scale of small source fluctuations $D$ and the system fluctuations $D_d$ in large. The parameter $D$ depends on the type of a substance and temperature. Small-scale high-frequency fluctuations attributed to a motion near one of the equilibrium positions can be estimated based on the solution for a harmonic Brownian oscillator

$$\frac{\partial \eta}{\partial t} = -2\eta + \sqrt{2D}\xi(t) \tag{17}$$

with spectrum

$$S_0(\omega) = \frac{2D}{1+\omega^2}. \tag{18}$$

The value $D$ sets by the average magnitude of fluctuations $\langle \eta^2 \rangle$. Due to the potential symmetry, the high-frequency spectrums (18), averaged over the time spent in each well, is identical for both wells. Finally, the complete fluctuation spectrum is the sum of the low-frequency and high-frequency components: $S(\omega) = S_d(\omega) + S_0(\omega)$.

## 4. Results of numerical simulation

The equation of motion (13) coincides with the Duffing equation for overdamped nonlinear oscillator under the action of a random force [32]. The correlation function for a position of this system may have different spectra, depending on the chosen parameters of the oscillator and noise [33]. In particular, asymmetric distributions far from the Lorentz form are generated. However, the earlier power spectrum calculations [34, 35] did not clarify the nature of the high-frequency asymptotic behavior of fluctuations. To obtain more complete description, we reproduce the fluctuation dynamics by computer simulation of the solution ensemble to the Langevin equation (13) [36]. Figure 2 shows a sample result for the time series of the induced fluctuations. Its profile demonstrates the two-scale dynamics of the nonlinear oscillator in the critical region and is consistent with the analog dependence measured for this oscillator [26]. The dynamics shows high-frequency chaotic oscillations near equilibrium positions, which are alternated by spasmodic transitions from one potential well to another. It is important to note that the experimental dependencies of the heat flow on time during premelting contain qualitatively similar two-scale components, but switching are not so abrupt [14].

Experimental data for the time dependence of random fluctuations are not repeated from one case to another and cannot be directly compared with the random realizations in numerical simulation. A relevant characteristic of this stochastic process is the power spectrum $S(\omega)$, which we obtained for a bistable oscillator by numerical simulation of a solution ensemble. On the modeling time interval, large-scale oscillations were generated with an average number of such oscillations equal to 40. Averaging was carried out over an ensemble of 300 realizations.

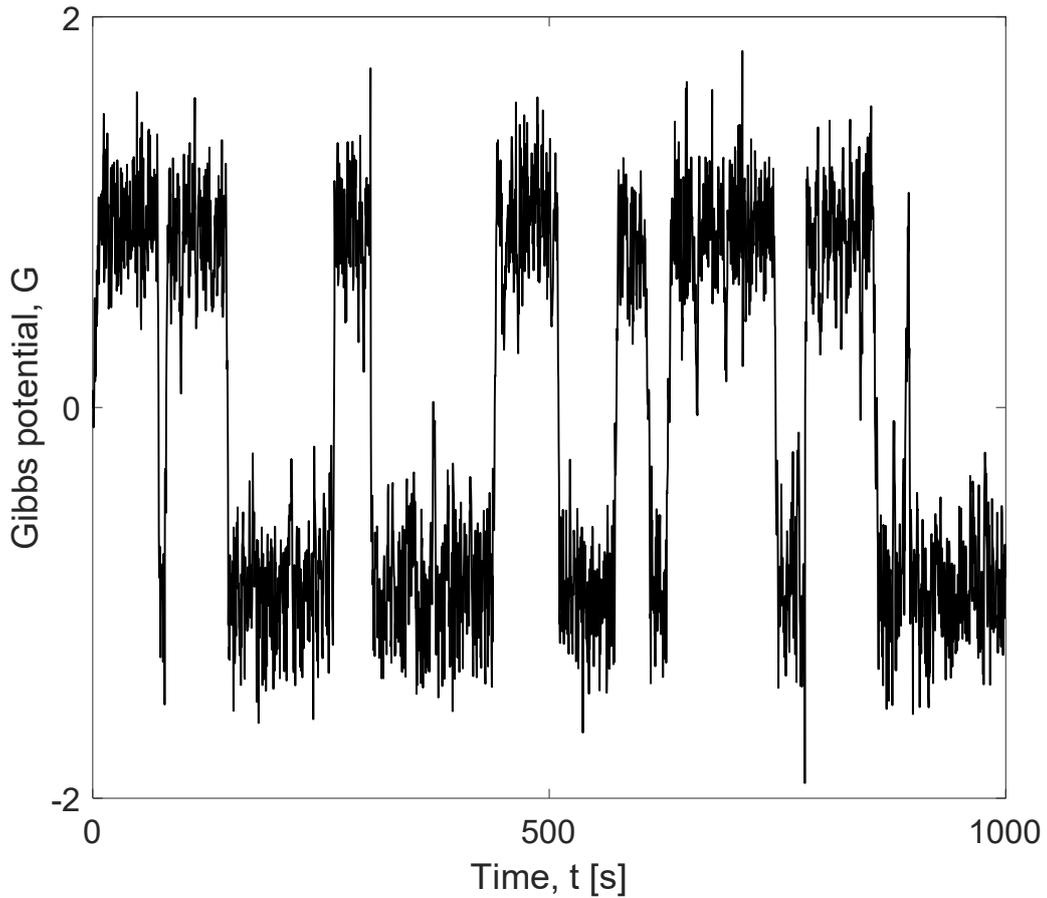

Figure 2. Example of numerical simulation of the dynamics of a bistable oscillator under the influence of random white noise.

The obtained results demonstrate robustness to increasing the length of the time interval and statistical sample size. The spectrum for the parameter value $D = 0.07$ is shown in Figure 3 in comparison with the spectrum calculated in the two-scale analytical model. It is robust with respect to the small disturbances in the potential symmetry and is sensitive to variations in a magnitude of $D$.

A comparison of theoretical and experimental dependences for Cu shows that both the analytical and numerical models give spectra consisting of two domains with different rate of decreasing power spectrum with increasing frequency. The result of computer modeling shows the dependence is characterized by two quasilinear slopes in the double-logarithmic scales, being founded earlier analyzing experimental data [15]. At the same time, our simplified analytical model does not reproduce such asymptotic behavior. A more advanced analytical model based on the methods of quantum field theory in the Kraichnann-Wild approximation [37] gives a spectrum structure more or less similar to experimental one, but does not clarify its nature. We note that both our theoretical models proceed from the single nonlinear Langevin equation and give qualitatively similar conclusions, but with much better agreement between result for direct numerical simulation and experiment.

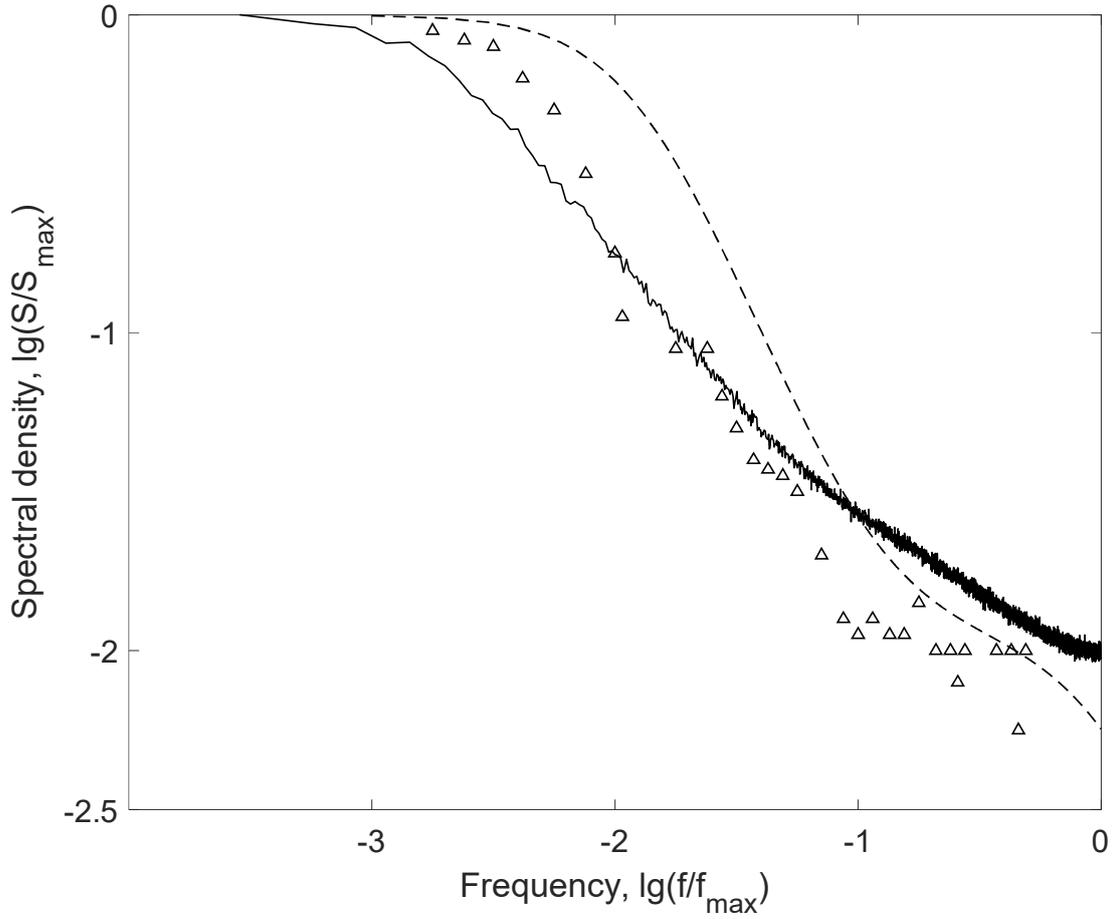

Figure 3. The dependence of the relative spectral density of fluctuations $\lg(S/S_{max})$ for a Cu sample on the relative frequency $\lg(f/f_{max})$, where $S_{max}$ and $f_{max}$ are the maximum values. The solid line shows the result of direct numerical simulation. The dashed line demonstrates the spectrum calculated with the two-scale analytical model. The experimental data [15] are indicated by triangles.

Thermal fluctuation coefficient $D$ for the medium in the premelting condition can be estimated using linear equation [38] for small fluctuations near equilibrium as

$$D = \frac{4\beta k_B T}{V\alpha^2}, \qquad (19)$$

where $\beta$ is the coefficient of isothermal compressibility, $k_B$ is the Boltzmann constant, $V$ is the volume of the premelting microstructure, $\alpha$ is the relative density variation during the melting. For Cu samples, we have parameter values $\beta = 0.0136\,\text{GPa}^{-1}$, $\alpha = 0.04$, [39], and a numerical estimation shows that at the premelting temperature $T = 1293\,\text{K}$ [14], fluctuations at the level obtained when modeling the observed premelting spectrum occur in structures with characteristic sizes of several nanometers, depending on a grain geometrical form. This indicates the decisevly important role of polycrystalline copper grains without dislocations [40] in the process of premelting and restrictions of uniform mean field model upon attempt a complete explanation of the experimentally observed fluctuations.

## 5. Conclusions

Based on the results of numerical simulation for a mean field model, we can formulate some conclusions about the nature and mechanism of premelting fluctuations in polycrystals. The high-frequency part of these fluctuations is associated with chaotic density fluctuations in each of the coexisting phases. Low-frequency fluctuations are associated with more rare transitions from one phase to another. The description of both types of fluctuations in the main field theory is reduced to solving the Langevin equation for a nonlinear bistable oscillator. We developed analytical two-scale model of the premelting fluctuations by separation of random motions into two types: oscillations near equilibrium positions and jumps between them, the rate of which is described by Kramers theory. This simplified model qualitatively correct predicts the general form of the fluctuation spectrum and the relative contribution of low and high frequencies. Our numerical simulation, like the analytical model, depends on a single parameter that is the intensity of thermal vibrations of the medium. The spectrum of fluctuations obtained by numerically solving the stochastic equation is better agreed with the experimental data and reproduced two spectrum domains with different power spectrum slopes. With our theoretical model, we supported the assumption [15] about existing two levels of a unified physical process of isothermal fluctuations in premelting. The remaining discrepancy between the theory and experiment can be attributed to details remained outside of the developed approach. Evaluations show that fitting the model with experimental data becomes adequate when assuming nanocrystalline structure of the substance.

To derive a more comprehensive dynamical model of premelting, it is important to consider the polycrystalline structure of the medium in more detail including statistical distribution of grains on size, as well as the thermal formation of defects and clusters [41]. In part, the effects of melting heterogeneity can be incorporated in the model in the form of Landau-Khalatnikov gradient potential [38]. One of additional and significant assumption in our model was the Markovian type of relaxation process, deviation from which can also affect the fluctuations. In the further investigations, it would be interesting, beside of all, to concentrate on the premelting control by forced acoustic oscillations, including stochastic resonance phenomenon allowing completely modify fluctuation spectrum.

**Acknowledgments**
The author is gratefully acknowledged to L.A. Bityutskaya for interpretation of experimental data.

**Disclosure statement**
No potential conflict of interest was reported by the author.

**Funding**
This work was supported by internal funding of the Voronezh State Technical University without attracting additional grants or other sources of funding.

**ORCID**
Pavel Golovinski ID https://orcid.org/0000-0002-7527-0297.